# Description of ResFrac automated history matching and optimization workflow

Charles A. Kang, charles@resfrac.com

Mark W. McClure, mark@resfrac.com

Somasekhar Reddy, soma@resfrac.com

ResFrac Corporation

Nov. 25, 2021

## Introduction

This document describes the functioning of the automated history matching and optimization workflow implemented with the ResFrac hydraulic fracturing and reservoir simulator. The purpose of the automated history matching and optimization workflow is enable automated calibration of simulator models to observed data and application to forward optimization. The automated workflow works by solving a formal mathematical optimization problem to minimize misfit with observations from any point in the lifecycle of a hydraulically fractured reservoir, or to maximize a quantity of interest associated such a reservoir, such as net present value. Objective function evaluations in the optimization problem consist of runs of the ResFrac simulator, which is a physics-based model of hydraulic fracturing and reservoir phenomena. The workflow employs a proxy model to improve computational speed and applies experimental design and Bayesian sampling techniques to generate points with which to train the proxy model. This document also provides an overview of the software infrastructure developed to support the automated workflow.

## ResFrac simulator

The ResFrac hydraulic fracturing and reservoir simulator is at the core of each function evaluation. The simulator fully couples a three-dimensional planar hydraulic fracture model with geomechanics and reservoir flow models. The fully integrated nature of the ResFrac simulator makes it uniquely well-suited for modeling unconventional resource development, as ResFrac simulations encompass the whole life cycle of a well. As such, it straightforward to use the ResFrac simulator to develop history matching workflows that match observations obtained both during the fracturing phase (such as fracture length, inferred from datasets such as microseismic and offset well observations), and during the production phase (such as fluid production histories). Details on the simulator are available in previously published literature [1, 2].

## History matching and optimization workflow overview

We consider the following mathematical optimization problem:

$$\operatorname*{argmin}_{\mathbf{z}} \hat{j}_{\text{sim}}(\mathbf{z}), \text{subject to } \mathbf{z} \in \Omega,$$

*Equation 1*

where $\hat{j}_{\text{sim}}$ is an objective function, **z** is a vector of input parameters representing uncertain quantities that are varied in the history match (such as geological properties), and $\Omega$ is the search domain. In history matching, the objective function represents the degree of misfit between simulator outputs and observed data, while in optimization, the objective function typically represents some quantity of economic interest such as NPV or cumulative oil production (the sign on the objective function is flipped as needed). The simulator input space is parameterized and scaled such that **z** is relatively low dimension ($n_{dim} \sim 1 - 10$) and the individual entries in **z** are bounded between $-\mathbf{1}$ and $+\mathbf{1}$. In principle, Equation 1 could be solved directly by applying an optimization algorithm. However, because each evaluation of $\hat{j}_{\text{sim,reg}}$ entails a computationally costly simulator run, we do not directly use the simulator in an optimization loop.

Instead, we approximate simulator runs using a fast-running proxy model, $\hat{j}_{\text{proxy}}(\mathbf{z}) \approx \hat{j}_{\text{sim}}(\mathbf{z})$. Within the context of the optimization problem considered here, the proxy-model-based objective function has the same input and output domains as the simulator-based objective function. It is important to note that the proxy model is not a general-purpose substitute for the simulator, however, as it only represents the subset of simulator outputs that are relevant to the specific optimization problem. Applying the proxy model approximation, we then consider following the mathematical optimization problem:

$$\underset{\mathbf{z}}{\text{argmin}}\, \hat{j}_{\text{proxy}}(\mathbf{z}), \text{subject to } \mathbf{z} \in \Omega.$$

*Equation 2*

To ensure that the proxy model accurately captures the behavior of the simulator in the regions of interest, we employ an iterative sample-train-optimize procedure illustrated in. In each iteration of this procedure we train a proxy model, optimize on the proxy model, and then identify additional points to run in the next iteration. The sample of points is selected such that the proxy model is refined in areas of interest to the optimization (i.e., those regions where with objective function value is close to the optimum).

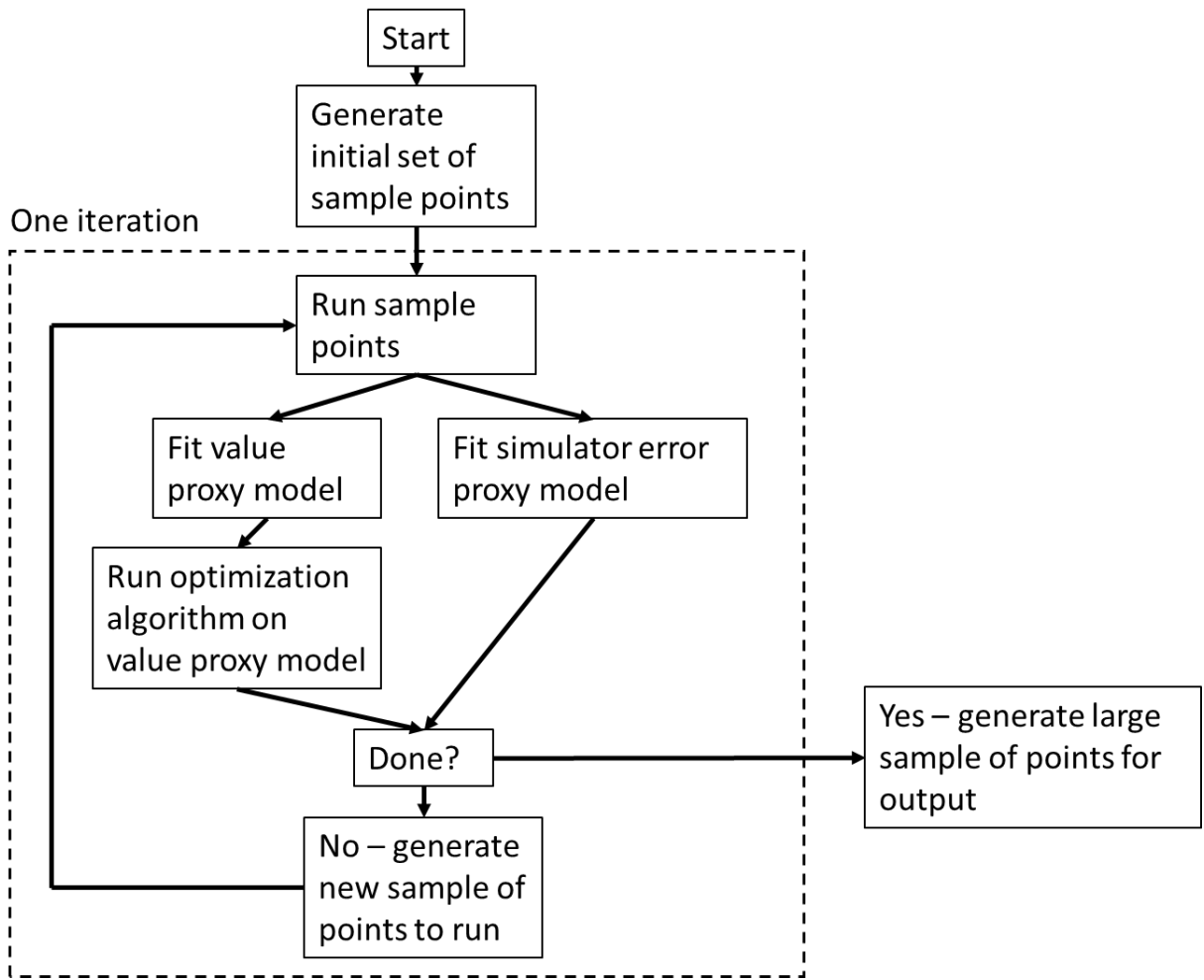

*Figure 1 Automated history matching workflow*

Each iteration consists of the following steps. First, a sample of points from the search domain is generated and run in the simulator. After the simulation points in the iteration are completed, a proxy model (the 'value' proxy model) is trained for the objective function. Additionally, a second proxy model (the 'simulator error' proxy model) representing uncertainty in simulator outputs is trained. Next, a sampling procedure that uses the value and simulator error proxy models is applied to identify a new set of points with objective function value comparable to or better than that of the best point identified so far. Finally, an optimization algorithm is run on the value proxy model. The 'best point' identified by the optimization algorithm, along with the set of points identified in the sampling procedure, are then used as the set of points to run in the next iteration.

A Bayesian interpretation of the automated history matching workflow is that the workflow starts with an uninformative prior and in each iteration adds to the state of information by expanding the sample of simulations. The additional simulations that are run in each iteration are selected using a heuristic be in areas that have or potentially have favorable objective function value, so that the state of information is most enhanced in regions of interest (i.e., those that could potentially have favorable objective value). At the end of the workflow, a sampling procedure can be applied to generate a large set of points that can be marginalized to provide estimated distributions of input parameters that have favorable objective function values (i.e., that minimize misfit with observed data). This can be used to, for example, determine whether a given history matching or optimization problem has a unique or nonunique solution, and assess the extent of uncertainty that remains after matching observed data.

## Objective functions

In this document we describe three types of objective functions: time series misfit, goal value misfit, and net present value (NPV). The misfit functions are minimization objectives employed together in the model fitting examples in this work. NPV is a maximization objective that is optimized after the model fitting step.

Time series misfit objective functions are calculated as the root mean square difference between simulation output and observations over the range in time of available data:

$$\hat{y}_{\text{time series misfit}} = \left[\frac{1}{n}\sum_{i=1}^{n}(\hat{y}_{i,\text{sim}} - y_{i,\text{obs}})^2\right]^{1/2},$$

*Equation 3*

where $n$ is the number of evenly spaced sample points (we use 10,000), $y_{i,\text{sim}}$ is the simulated data, and $y_{i,\text{obs}}$ is the observed data. The quantities $y_{i,\text{sim}}$ and $y_{i,\text{obs}}$ are linearly interpolated when the sample time does not coincide with a simulator output time or observation time.

Goal value misfit objective functions are calculated as the absolute value of the difference between simulation output and observations at a specific point in time:

$$\hat{y}_{\text{goal value misfit}} = |\hat{y}_{\text{sim}} - y_{\text{obs}}|,$$

*Equation 4*

where $\hat{y}_{\text{sim}}$ is the simulation output value and $y_{\text{obs}}$ is the goal value.

NPV objective functions are calculated as the discounted cash flows (revenues less expenses) of a simulation:

$$\hat{y}_{\text{NPV}} = \sum_{t=1}^{T}\frac{\hat{c}_t}{(1+r)^t},$$

*Equation 5*

where $T$ is the number of years, $\hat{c}_t$ is the predicted net cash flow in year $t$, and $r$ is the discount rate. Because the optimization problem is formulated as a minimization in Equation 1, we flip the sign so that NPV is maximized (i.e., we minimize $-\hat{y}_{\text{NPV}}$). Other economic objectives could include quantities such as cumulative oil and gas production.

Additionally, any other time series output from the simulator could be selected as a general maximization or minimization objective, either over a range in time (minimum, average, or maximum over that range), or at a single point in time.

## Objective function scaling

Objective functions are scaled such that the objective function values are on order 1. The scaled objective function value is calculated from the raw objective function value based on a scaling constant:

$j = \frac{y}{s}$, where $j$ is the scaled objective function value, $y$ is the raw objective function value, and $s$ is the scaling constant for that objective. Similarly, the proxy model standard error estimate is calculated as $\hat{\sigma} = \frac{\hat{\sigma}_{\text{raw}}}{s}$, where $\hat{\sigma}$ is the scaled standard error estimate, $\hat{\sigma}_{\text{raw}}$ is the unscaled standard error estimate, and $s$ is scaling constant (the same scaling constant is used for the objective and for its corresponding standard error estimate). For goal value objectives, the goal value is used as the scaling constant. For misfit objectives, the scaling constant is calculated as the average of the time-weighted median value in the time series, and the time-weighted standard deviation of values in the time series.

## Objective function lumping

In scenarios with more than one objective function, we combine the multiple objectives together into a single objective. Multiple objectives are lumped by taking a linear combination of the scaled objective function values:

$$\hat{j}_{\text{lumped}} = \sum_i a_i \hat{j}_i,$$

*Equation 6*

where $a_i$ represents the weight for the $i$th objective function ($a_i > 0$ for all $i$, and the weights are normalized such that $\sum_i a_i = 1$), and $\hat{j}_i$ represents the proxy model's prediction for the $i$th objective function value. We calculate the lumped estimate of standard error with the assumption that the estimates of standard error for the individual objectives are Gaussian and independent, i.e., by taking a weighted average of the variances:

$$\hat{\sigma}_{\text{lumped}} = \left( \sum_i a_i \hat{\sigma}_i^2 \right)^{1/2},$$

*Equation 7*

where $\hat{\sigma}_i$ represents the estimate of the standard error of the prediction for the $i$th objective.

## Regularization

We apply regularization to bias the optimization algorithm and sampling procedure toward a certain value. This makes the solution more well behaved because it applies a weak convexity to the search space. Regularization can be interpreted as applying a Bayesian prior. The regularized objective function is calculated as

$$\hat{j}_{\text{tot}} = (1-\lambda)\hat{j}_{\text{unreg}} + \lambda j_{\text{reg}},$$

*Equation 8*

where $\lambda \in [0,1)$ is the regularization weight constant. We employ the $L_2$-regularization, wherein the regularization function has the form $j_{\text{reg}} = \|\mathbf{z} - \mathbf{z}_{\text{regcenter}}\|_2$. This selection of functional form corresponds to a Gaussian prior. Alternative functional forms for regularization such as $L_1$, $L_\infty$, $L_p$, or elastic net, corresponding to different priors, could also be used. The standard error estimate for the regularized objective function is calculated as $\hat{\sigma}_{\text{tot}} = (1-\lambda)\hat{\sigma}_{\text{unreg}}$, where $\hat{\sigma}_{\text{unreg}}$ is the unregularized standard error.

By default, we apply a center of regularization of $\mathbf{z}_{\text{regcenter}} = \mathbf{0}$, i.e., centered in the search domain, but the user can select alternative coordinates for the regularization center. Similarly, the regularization weight defaults to $\lambda = 0.01$, meaning that regularization contributes one percent of the lumped scaled objective function value, but the user can specify a different value for this quantity.

### Optimization algorithm

We use the COBYLA algorithm as implemented in the SciPy library [3] to solve all optimization problems in this workflow. The COBYLA algorithm is a direct-search algorithm that works by solving a sequence of linear approximations to the nonlinear problem [4]. Within each optimization run, to reduce the impact of local minima on the search process, we employ a multi-start technique in which the optimization algorithm is run multiple times with different initial guesses points. The set of initial guesses includes the best previously run simulation point and a set of pseudorandom points drawn uniformly from the input domain. Note that the sets of initial guesses used in the optimization algorithm are distinct from the sets of sample points in the proxy model fitting workflows described below; the sets of initial guesses are always generated using the same procedure (superset of the best previously run simulation point and set of pseudorandom points). We take the best point resulting from any of the optimization runs as the result of the optimization algorithm.

### Proxy model fitting

The procedure for fitting a proxy model consists of training with a set of simulation inputs and outputs from simulation runs, $\{(\mathbf{z}_i, y_i) \forall i\}$, where $\mathbf{z}_i$ represents the coordinates of the $i$th simulation point and $y_i$ is the simulator output corresponding that point. The resulting proxy model can evaluate any point in the input space to provide a prediction for the simulator output and a prediction of the standard error of the proxy model at that point. For multi-objective problems, an independent proxy model is constructed for each objective. We employ Kriging as implemented by the Python Surrogate Modeling Toolbox [5]. Alternatively, any proxy model that can provide an estimate of its own model error could be used in this workflow, with the caveat that the proxy model should be smooth if the optimization algorithm applied is derivative-based (the COBYLA algorithm that we apply does not use derivatives).

### Simulator uncertainty

Outputs from the ResFrac simulator are inherently uncertain, with sources of uncertainty including incomplete knowledge about the state of the subsurface, incomplete understanding of physical processes, measurement, and random noise. Some of this uncertainty can be represented within the simulator. For example, ResFrac simulations can incorporate a pseudorandom variation in fracture toughness to account for variation in subsurface properties; thus, ResFrac simulations could be employed in an ensemble-based workflow employing multiple different random seeds. We do not presently consider an ensemble-based workflow, though this would be a natural extension of this workflow. However, we do attempt to account for some aspects of inherent simulator uncertainty through a crossvalidation-based procedure.

We conceptualize simulation outputs from individual runs as comprising an average simulator response combined with an error term that is a random variable whose behavior varies over the input space:

$$\hat{y}_{\text{sim}}(\mathbf{z}) = y_{\text{avg}}(\mathbf{z}) + \omega_{\text{err}}(\mathbf{z}),$$

*Equation 9*

where $\hat{y}_{\text{sim}}(\mathbf{z})$ represents the simulator output, $y_{\text{avg}}(\mathbf{z})$ represents the average behavior of the simulator, and $\omega_{\text{err}}(\mathbf{z})$ represents a draw from the error distribution associated with the simulation run.

We construct a model to predict the simulator error term, $\omega_{\text{err}}(\mathbf{z})$, by applying leave-one-out crossvalidation in the following manner. (Note: other crossvalidation techniques, such as k-fold crossvalidation, could be applied instead of leave-one-out crossvalidation in this procedure, with similar results.) For each simulation point in the sample set, we construct the 'value' proxy model using all points except for the point being considered. We then evaluate the value proxy at that point. The resulting proxy model prediction value is treated as a synthetic rerun of the simulator. The proxy model's self-generated error does not play a role in this procedure, as the self-generated error estimate represents the proxy model's estimate of its prediction error relative to the simulator results, whereas in this procedure the goal is to estimate error resulting from random noise in simulator outputs. We treat the leave-one-out proxy model's prediction at the left-out point as an independent observation, and in so doing we have two independent simulation draws at the same point in the input space: $\hat{y}_{\text{sim}}$ from the simulator run, and $\hat{y}_{\text{proxy}}$ from the leave-one-out value proxy model evaluation. We assume that simulator error follows a Gaussian distribution with mean zero, $\omega_{\text{err}}(\mathbf{z}) \sim N\left(\mathbf{0}, \sigma^2_{\text{sim}}(\mathbf{z})\right)$, and so estimate the simulator standard error at the left-out point using the formula for estimating the standard error with a sample size of two observations, $\hat{\sigma}_{\text{sim}} = \frac{1}{\sqrt{2}}|\hat{y}_{\text{sim}} - \hat{y}_{\text{proxy}}|$.

Applying the above procedure for all points in the sample set yields an estimate of the standard error of the simulator at all simulation points in the sample set. Finally, we fit a 'simulator error' proxy model, $\hat{\sigma}_{\text{sim}}(\mathbf{z})$, to this dataset. We use the same fitting procedure and algorithm for the simulator error proxy model as for the value proxy model. The simulator error proxy model provides an estimate of the standard error associated with the simulator at any point in the input domain.

### Sample generation for the first iteration

In the first iteration, we start with no information about the shape of the objective function surface, so we employ experimental design sampling techniques, such as described in [6] and [7], to generate a space-filling set of points. We employ different experimental designs, depending upon the number of dimensions in the input domain. For $n_{\text{dim}} \leq 3$, we use a full factorial design is applied; for $4 \leq n_{\text{dim}} \leq 5$, we use a centered composite design; and for $n_{\text{dim}} \geq 6$, we use a superset of $k n_{\text{dim}}$ points (where $k = 10$) including the center point, the one-at-a-time extreme values along each dimension, and a pseudorandom Latin Hypercube sample. The experimental designs are implemented using the pyDOE2 library [8].

### Sample generation for later iterations

In later iterations, we make use of the fact that we now have information about the search space when selecting additional points to run. We apply a procedure based on a rejection sampler. The purpose of this procedure is to produce points that will increase the accuracy of the proxy model in regions of the input space that potentially have favorable objective function values – i.e., in regions of the input space where the objective function value is predicted to be favorable, or in regions where the uncertainty is high and thus potentially may host favorable objective function values. In the sampling procedure, we (1) identify the best objective value achieved so far and a quantification of its associated uncertainty, (2) generate a candidate point, (3) accept or reject the candidate based on a how its estimated objective value and uncertainty compare with that of the best achieved objective value and uncertainty.

In the first step, we evaluate the regularized lumped objective value for all previously run simulations, and we record the value for the best point, $\hat{j}_{\text{best}}$. We use the model error proxy model to estimate the simulator standard error for this point, $\hat{\sigma}_{\text{best}}$.

In the second and third steps, candidate points are drawn and tested for acceptance according to the following procedure.

   a. A candidate point is drawn from a uniform distribution on the input domain.
   b. The value proxy model is evaluated at the coordinates of the point, yielding an estimate of the value, $\hat{j}_{\text{cand}}$, and of proxy model standard error, $\hat{\sigma}_{\text{proxy,cand}}$ (recall that the value proxy model produces estimates for both the value and the standard error of the value estimate).
   c. The simulator error proxy model is evaluated at the point to produce an estimate of simulator error, $\hat{\sigma}_{\text{sim,cand}}$ (recall that this is not the same as $\hat{\sigma}_{\text{proxy,cand}}$; $\hat{\sigma}_{\text{sim,cand}}$ represents uncertainty in the simulator output, while $\hat{\sigma}_{\text{proxy,cand}}$ represents uncertainty in the proxy model's prediction of simulator output).
   d. The overall combined error estimate of the difference between the lumped objective value at the point and at the best point is calculated, applying the assumption that errors for the candidate point and at the best point are Gaussian and independent, as $\hat{\sigma}_{\text{tot}} = \left(\hat{\sigma}_{\text{cand}}^2 + \hat{\sigma}_{\text{best}}^2\right)^{1/2}$, where $\hat{\sigma}_{\text{cand}} = \left(\hat{\sigma}_{\text{sim,cand}}^2 + \hat{\sigma}_{\text{proxy,cand}}^2\right)^{1/2}$ represents the combined simulator and proxy model error estimate of the proxy model evaluation of the candidate point, and $\hat{\sigma}_{\text{best}}$ is the simulator error estimate at the best point (since the best point is an actual simulation, there is no proxy model error contribution to the estimate of error for the best point).
   e. The lumped objective value for candidate point is then compared to that for the best point by calculating a score for the candidate, $\hat{\chi}_{\text{cand}} = \frac{\hat{j}_{\text{cand}} - \hat{j}_{\text{best}}}{\hat{\sigma}_{\text{tot}}}$.
   f. The candidate point is accepted or rejected by comparing its score with that of a random draw $\chi \sim N(0,1)$. If $\hat{\chi}_{\text{cand}} \leq \chi$, the point is accepted, and otherwise the candidate is rejected.

We repeat the candidate generation-acceptance/rejection steps until a desired number of accepted points is obtained. The goal number of accepted points in each iteration is calculated using a heuristic, $n_{\text{goal}} = \min(10 n_{\text{dim}}, \lceil 5 \times 1.5^{n_{\text{dim}}} \rceil)$.

For certain problems, such as those in which the region of the search space that have favorable objective values (relative to the best observed objective value) is small relative to the size of the domain, the rejection sampling procedure described above can have a very low acceptance rate. If this is the case, then an alternative sampling procedure would need to be applied. The alternative sampling procedure could be substituted within this workflow without affecting the rest of the workflow. This potential issue is typically not relevant for problems with relatively low dimension (in our experimentation with history matches of up to 10 input parameters, the rejection sampler has not posed a serious problem).

In addition to being used to generate points to run for the subsequent iteration, the infill sampling procedure is also applied to generate large samples of points (we use 1,000 points) at the end of each iteration. These large samples of points are used in outputting and visualization to represent the state of information for visualization.

## Software infrastructure

We have implemented the automated history matching and optimization workflow described in this document in commercial software. A desktop-based graphical user interface enables users to set up and run history matching and optimization problems, and to download results as the workflow runs. The simulation orchestration software, which is at the core of the automated workflow, runs on a public cloud service. This orchestration software makes use of a command-line interface that manages simulation data. Simulations are also run in a public cloud service, and ResFrac server orchestration software autonomously allocates and deallocates computational resources as needed to run simulations efficiently.